\documentclass[
    reprint,
    superscriptaddress,
    amsmath,
    amssymb,
    aps,
    prx,
    longbibliography
]{revtex4-1}

\usepackage{physics}

\usepackage{graphicx}
\usepackage{dcolumn}
\usepackage{bm}
\usepackage[colorlinks=true,urlcolor=blue,linkcolor=blue,citecolor=blue]{hyperref}
\usepackage[normalem]{ulem}
\usepackage{tikz}
\usetikzlibrary{quantikz}
\usepackage{amsmath}
\usepackage{amssymb}
\usepackage{bbold}

\newcommand{\customref}[2]{\hyperref[#1]{\ref*{#1}#2}}

\usepackage{xcolor}

\definecolor{Ured}{HTML}{cc0000}
\definecolor{Ublue}{HTML}{1f65cf}
\definecolor{Ugreen}{HTML}{198a11}

\newcommand{\pf}{p_{\rm f} }
\newcommand{\pme}{p_{\rm m} }
\newcommand{\rhoavg}{\overline{\rho}}

\begin{document}

\title{Entanglement and Absorbing-State Transitions in Interactive Quantum Dynamics}

\author{Nicholas O'Dea}
\affiliation{Department of Physics, Stanford University, Stanford, CA 94305, USA}

\author{Alan Morningstar}
\affiliation{Department of Physics, Stanford University, Stanford, CA 94305, USA}

\author{Sarang Gopalakrishnan}
\affiliation{Department of Electrical Engineering, Princeton University, Princeton, NJ 08544, USA}

\author{Vedika Khemani}
\affiliation{Department of Physics, Stanford University, Stanford, CA 94305, USA}

\begin{abstract}

Nascent quantum computers motivate the exploration of quantum many-body systems in nontraditional scenarios. For example, it has become natural to explore the dynamics of systems evolving under both unitary evolution and measurement. Such systems can undergo dynamical phase transitions in the  entanglement properties of quantum trajectories \emph{conditional} on the measurement outcomes.
Here, we explore dynamics in which one attempts to (locally) use those measurement outcomes to steer the system toward a target state, and we study the resulting phase diagram as a function of the measurement and feedback rates. Steering succeeds when the measurement and feedback rates exceed a threshold, yielding an absorbing-state transition in the trajectory-averaged density matrix. We argue that the absorbing-state transition generally occurs at different critical parameters from the entanglement transition in individual trajectories and has distinct critical properties. The efficacy of steering depends on the nature of the target state: in particular, for local dynamics targeting long-range correlated states, steering is necessarily slow and the entanglement and steering transitions are well separated in parameter space.

\end{abstract}

\maketitle

%%%% INTRO %%%%

%\emph{\textbf{Introduction---}} 
Phases and criticality in quantum many-body dynamics are topics of fundamental interest. For unitary dynamics, various questions have been explored, such as the growth of entanglement~\cite{calabrese2005evolution, kim2013ballistic, nahum_quantum_2017}, the emergence of hydrodynamics~\cite{Khemani-Huse2018_emergence, Rakovszky-vonKeyserlingk2018_diffusive,Brown-Bakr2019_bad}, or the effects of disorder~\cite{Nandkishore-Huse2015_review,abanin_colloquium_2019,Alet2018_review}.
Recently, motivated by advances in quantum computing hardware, ``monitored" quantum systems subject to repeated local measurements have also come under intense investigation~\cite{potter_entanglement_2021,Fisher_random_quantum_circuits}. These systems were shown to display the surprising phenomenon of entanglement phase transitions as a function of measurement rate~\cite{li_quantum_2018,skinner_measurement-induced_2019,choi_quantum_2020,gullans_dynamical_2020,bao_theory_2020,li_measurement-driven_2019,Jian-Ludwig2020_measurement_criticality,ippoliti_entanglement_2021,Agrawal_2022}. 
Such phase transitions are only visible in the properties of \emph{individual} quantum trajectories~\cite{Plenio_quantum_jump} corresponding to specific sequences of measurement outcomes,
so they bear a prohibitive post-selection cost that is exponential in the space-time volume (with exceptions in Clifford dynamics~\cite{Noel_measurement_2022,Li_crossentropy}, spacetime-dual dynamics~\cite{Ippoliti_postselection, ippoliti_fractal_2022, Lu_2021}, or replacing measurements with swaps into an environment~\cite{weinstein2022scrambling}). 
Monitored dynamics can be enriched by using the measurement outcomes, say, for adaptively controlling the subsequent dynamics~\cite{Sierant-Pagano2022_dissipative_floquet,Cai2022_feedback, garratt2022measurements, Buchhold_preselection,Iadecola_dynamical,McGinley_feedback_dtc, Friedman_adaptive}. Error correction is one example of such \emph{interactive quantum dynamics} with a transition as a function of the error rate~\cite{nielsen2002quantum}. Understanding novel dynamical phenomena in interactive evolution is an active area of inquiry, especially topical in light of experimental advances in building devices capable of quantum control via measurements and feedback. 

\begin{figure}
\includegraphics[width=\columnwidth]{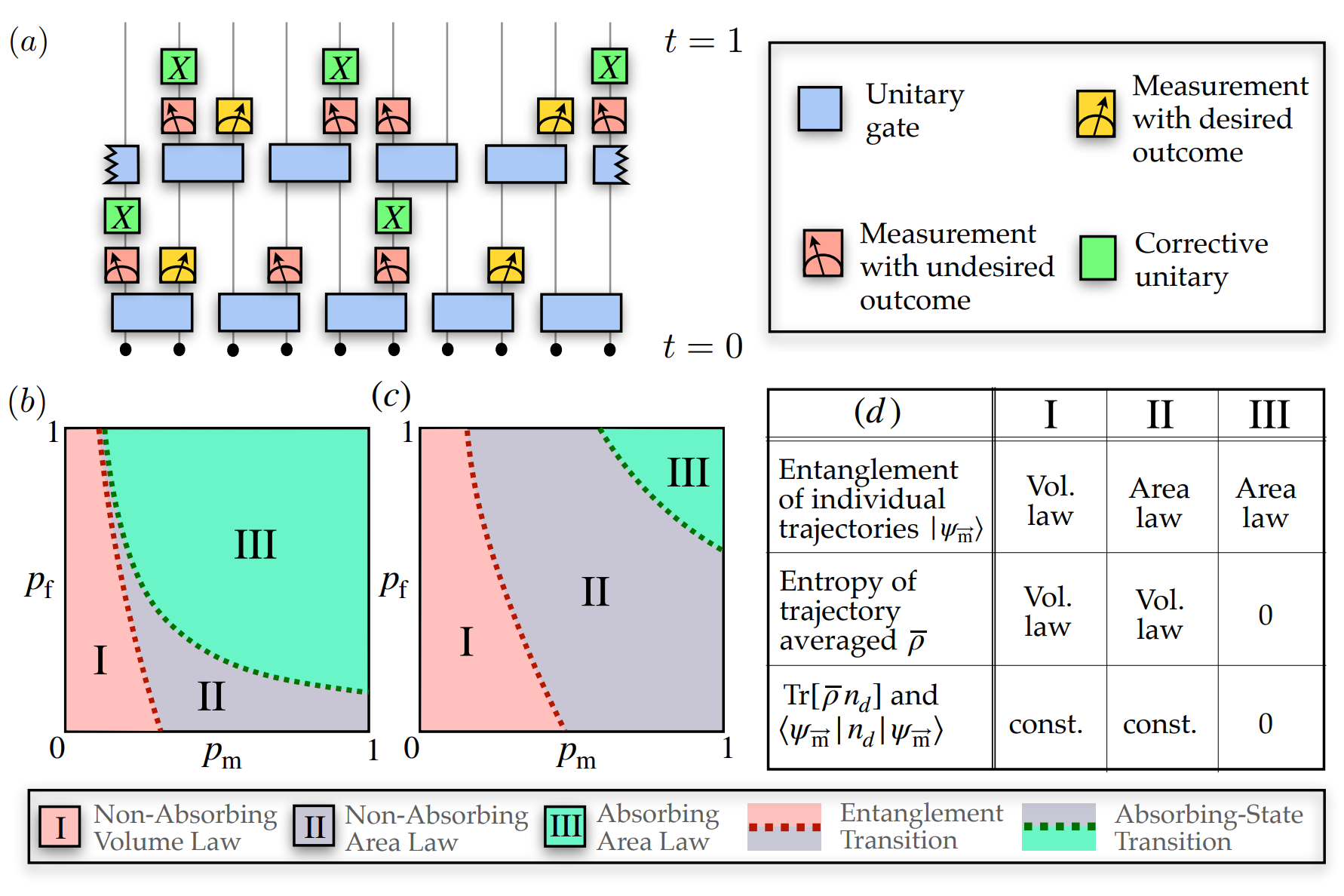}
   \caption{Monitored quantum circuits with feedback and schematic depictions of their associated phase diagrams. (a) A quantum circuit model with $\ket{\uparrow \uparrow \cdots \uparrow}$ as its target state (see text). 
   Sketch of the phase diagrams for interactive dynamics targeting (b) the all-up state and (c) an SPT cluster state. The two phase boundaries are close/widely separated near $\pf = 1$ in b/c respectively. The absorbing-state phase transition curves in (b,c) satisfy $\pme \pf = \rm{const}$. (d) A table of the properties of the three phases. 
\label{fig:circuit_and_phasediagram}}
\end{figure}

In this work, we consider interactive quantum dynamics with unitary evolution and measurements, where the measurement outcomes are used to apply local unitaries that \emph{steer} the system towards target states~\cite{Roy-Gefen2020_steering,Kraus-Zoller2008_preparation,Zhou-Lukin2021_preparation}.
The target states are absorbing states, so that the dynamics can evolve to the target state, but cannot leave it.  
We discuss steering both to trivial product states and to entangled symmetry-protected topological (SPT) states.
Our dynamics only uses local feedback
(in contrast to non-local classical communication), i.e., each feedback operation depends only on the measurement immediately preceding it at the same location. 
Therefore, the average density matrix dynamics is described by a time-independent local quantum channel. Unlike the quantum channels that occur in the standard (non-adaptive) measurement-induced transition, our channel is \emph{not} unital, so its steady state need not be the maximally mixed state. The channels we consider can in fact undergo an absorbing-state phase transition~\cite{Hinrichsen_2000, hinirchsen2008non, PhysRevLett.116.245701, PhysRevE.94.012138, PhysRevA.96.041602, lesanovsky2019non, PhysRevLett.123.100604, chertkov2022characterizing} separating an `absorbing phase' in which a pure zero-entropy target state is reached in a time at most polynomial in system size, from an `active phase' in which it is not.   

We study interactive quantum dynamics as a function of the measurement rate, $\pme$, and the fraction of measurements that are followed by feedback, $\pf$. This yields a phase diagram with two transitions: an entanglement phase transition in individual trajectories, driven by $\pme$, and an absorbing-state transition in the average density matrix, driven by the total rate of feedback events $\pme$~$\pf$ (Fig.~\ref{fig:circuit_and_phasediagram}). We provide numerical evidence and analytical arguments that these do not coincide in general. When we target a trivial polarized product state and use maximally efficient feedback, $\pf= 1$, the transitions occur at sufficiently close $\pme$ that we cannot numerically distinguish the locations of the critical points. Nevertheless, we find that quantities associated with the trajectory-averaged density matrix scale with directed-percolation critical exponents that are completely different from the exponents associated with the entanglement transition in individual trajectories. This finding is natural if one posits two separate transitions, but implies a drastic violation of one-parameter scaling otherwise. 

Errors above a target state with short-range correlations can be locally corrected, allowing for efficient local feedback. 
For target states with long-range correlations, such as a ferromagnet or a topologically ordered state, error correction requires pairing up domain walls or anyons over long distances, which requires long-range classical communication~\cite{Piroli_2021, hsieh_locc, verresen_locc_hierarchy}. Absent such communication, the error correction dynamics is described by a local quantum channel, which has a light cone. Steering from a product state to a long-range correlated state therefore takes a time that diverges with system size, even at $\pme = \pf = 1$ where all trajectories are area-law entangled. To explore these consequences of locality, we study dynamics targeting a symmetry-protected topological (SPT) state using both symmetric and symmetry-breaking operations. 
In the former case, but not the latter, the entanglement and absorbing-state transitions are well separated, consistent with the logic above.

%%%%%%%%%%%%%%%%% POLARIZED ABSORBING STATE %%%%%%%%%%%%%%%%%

\emph{\textbf{Polarized Absorbing State---}} We first consider a spin-1/2 model with the target state $\ket{\psi_t} = \ket{\uparrow\uparrow \uparrow \cdots \uparrow}$. The model comprises two-site nearest-neighbor unitary gates applied in a brickwork fashion in a 1D system of length $L$ with periodic boundary conditions. Each gate is block-diagonal and locally leaves the $\ket{\uparrow\uparrow}$ state invariant, but acts as a Haar-random unitary $U(3)$ in the 3$\times$3 block of states spanned by $\{\ket{\uparrow\downarrow}, \ket{\downarrow\uparrow}, \ket{\downarrow\downarrow}\}$.
The unitary dynamics leaves $\ket{\psi_t}$ invariant, but is otherwise chaotic. Each gate in the circuit is sampled independently. Unless otherwise stated, we always begin with the system in the $\ket{\downarrow \downarrow \cdots \downarrow}$ state. In between the unitary layers, we measure the Pauli $Z_i$ operator on each site with probability $\pme$. If the outcome is $+1$, locally corresponding to the $\ket{\uparrow}_i$ state, we do nothing. If the outcome is $-1$, a corrective unitary $X_i$ is applied with probability $\pf$  to steer the system towards $\ket{\psi_t}$ [Fig.~\ref{fig:circuit_and_phasediagram}(a)]. As a channel, this measure-and-correct operation is equivalent to a qubit reset, which is accessible on current hardware~\cite{chertkov2022characterizing}. A given quantum trajectory, $\ket{\psi_{\vec{m}}(t)}$ is labeled by the sequence of measurement outcomes, $\vec{m}$, encountered at each position until time $t$; the label also implicitly includes the action of feedback events conditioned on the outcomes. We denote the density matrix obtained by averaging over trajectories, feedback events, and choices of random circuits as $\overline{\rho}$. 

We describe the phase diagram of this model in Fig.~\ref{fig:circuit_and_phasediagram} using properties of trajectories and properties of $\overline{\rho}$ at times linear in system size.
When the total rate of feedback operations $\pme \pf$ exceeds a critical value $p_c \approx 0.1$, $\rhoavg$ approaches the absorbing state (phase III). Here, every trajectory approaches the same unentangled target state. 
When $\pme \pf < p_c$, $\rhoavg$ remains mixed up to times exponential in $L$. However, individual trajectories exhibit two phases, even at $\pf = 0$: a volume-law phase (phase I) and an area-law phase (phase II). In phase II, each trajectory stays active and visits \emph{different} area-law states, so $\rhoavg$ remains high-entropy. We will argue below that phase II intervenes between phase I and III, even when $\pf = 1$.

\begin{figure}
\centering
\includegraphics[width=\columnwidth]{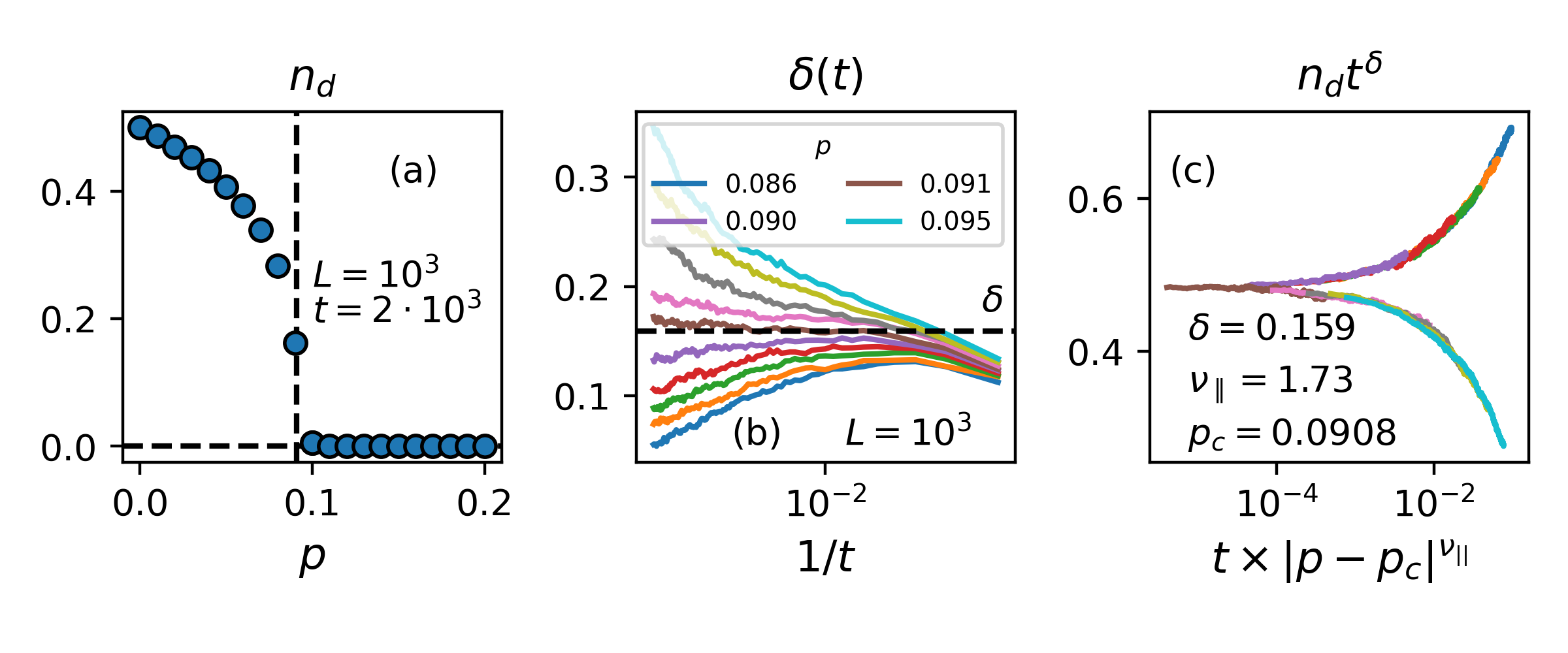}
\caption{Absorbing-state phase transition. (a) The sample-averaged defect density $n_d$ as a function of $p=\pme\pf$. Between $10^3$ - $2\cdot 10^3$ samples were averaged for each data point shown. Values were computed via the mapping to a classical stochastic model and initialized with random bitstrings. (b) The running estimate of critical exponent $\delta$ as a function of inverse time $1/t$. The critical $p_c$ is estimated to be the value of $p$ for which the curve remains constant as $1/t \to 0$, and an estimate for $\delta$ is that constant value. (c) Scaling collapse of $n_d$ data to the form of Eq.~\ref{eq:n_d_scaling}. The critical exponents are those of directed percolation. 
\label{fig:nd_classical}}
\end{figure}

\emph{Absorbing State Transition---} The absorbing-state transition is visible in $\rhoavg$, and observable via conventional expectation values. In~\cite{supp_mat}, we show that the diagonal elements of $\rhoavg$ evolve under the transfer matrix of a classical stochastic process, while the off-diagonal elements vanish after one complete time step.  This mapping gives us access to much larger system sizes and allows us to convincingly demonstrate that this absorbing-state transition is in the directed percolation universality class.
In the classical process,  the effect of the averaged measurements is to locally send $\downarrow$ to $\uparrow$ with probability $p_{\rm m} p_{\rm f}$, while the effect of the averaged unitaries is to mix the local configurations $\uparrow\downarrow$, $\downarrow\uparrow$, and $\uparrow\uparrow$ with equal probability $1/3$~\cite{supp_mat}. Note that the averaged dynamics are now described by only a single parameter $p \equiv p_{\rm m} p_{\rm f}$ describing the total rate of measurements with feedback.  This stochastic process has an all-up absorbing state, and undergoes an absorbing-state transition characterized by the behavior of ``defects", i.e., down-spins. The density of defects $$n_d = \frac{1}{L} \sum_i \frac{1-Z_i}{2}$$ serves as an order parameter that rapidly approaches zero (exponentially in time) in the absorbing phase, but reaches a long-lived nonzero value in the non-absorbing phase (for times exponential in $L$). In Fig.~\ref{fig:nd_classical}(a), we see that the density of defects is nonzero below $p_c\approx.1$ and vanishes above $p_c$. The curve $\pme \pf = p_c \approx .1$ defines the absorbing-state critical line depicted in Fig.~\ref{fig:circuit_and_phasediagram}(c). Note that a non-zero $n_d$ implies an extensive entropy for $\rhoavg$, with the entropy density being maximal at $n_d = 0.5$ and decreasing as $p_c$ is approached.  

For a classical stochastic process with a single absorbing state, a non-disordered transfer matrix, and no additional symmetries, the critical properties are expected to be in the directed percolation (DP) universality class. We confirm this expectation in Fig.~\ref{fig:nd_classical}(b,c), using techniques discussed in Ref.~\cite{Mendonca_2011}. We expect that $n_d$ satisfies the following critical scaling:
\begin{equation}
n_d(t, L) \sim t^{-\delta} \Phi\left((p-p_c)t^{1/\nu_{||}}, \frac{t^{1/z}}{L}\right).
\label{eq:n_d_scaling}
\end{equation}
where $\nu_\parallel = z \nu_\perp$ and $z$ is the dynamical scaling exponent. 
To probe the values of $\delta$ and $p_c$, we define a time-dependent estimate $\delta(t)$ so that $n_d(t) \sim 1 / t^{\delta(t)}$, and extract $\delta(t)$ from the quantity 
$\log_{10}[n_d(t) / n_d(10t)]$. We expect that at the critical point, and for $t^{1/z}/L$ sufficiently small, $\delta(t)$ is constant, while it vanishes in the non-absorbing phase and diverges in the absorbing phase. Fig.~\ref{fig:nd_classical}(b) shows a value of $\delta$  consistent with directed percolation's $\delta_{\rm DP} = .159$, and a critical probability  $p_c = .09085(5)$. We see scaling collapse of these curves in Fig.~\ref{fig:nd_classical}(c) using this $p_c$, $\delta_{\rm DP}$, and $\nu_{\rm DP ||}=1.73$.  Thus, as anticipated, our model's absorbing-state transition falls under the directed percolation universality class.

\emph{{Entanglement Phase Transition---}}
We now locate and characterize the entanglement phase transition via exact simulations of quantum trajectories for $L\le 24$.  
We first consider the cut $\pf=\pme$ through the phase diagram in Fig.~\ref{fig:circuit_and_phasediagram}(c); the feedback along this cut is weak enough that the entanglement phase transition is numerically well separated from the absorbing-state transition. We benchmark the entanglement transition against previous analyses of models without feedback. 
We use the tripartite quantum mutual information $I_3 = S_{Q_1} + S_{Q_2} + S_{Q_3} - S_{Q_1 \cup Q_2} - S_{Q_2 \cup Q_3} - S_{Q_3 \cup Q_1} + S_{Q_1 \cup Q_2 \cup Q_3}$, which was argued to be constant at the critical point and show a crossing~\cite{zabalo_critical_2020,gullans_dynamical_2020}. Here $S_A$ is the von Neumann entanglement entropy of subsystem $A$, and $Q_n$ is the $n^\mathrm{th}$ contiguous quarter of the system. 
We evaluate $I_3$ at $t = 2L$, anticipating that the entanglement transition will have dynamical exponent $z = 1$~\cite{li_measurement-driven_2019, Jian-Ludwig2020_measurement_criticality,li_conformal_2021}.
The crossing of $I_3$ yields the critical point $p_m^{c} = p_f^{c} \cong 0.130(5)$, which is well-separated from the absorbing-state phase transition at $\pme = \pf  = \sqrt{p_c} \cong \sqrt{0.09085} = 0.3014$ found above (Fig.~\ref{fig:nd_classical}).
We perform a scaling collapse near the critical point, and the critical exponent $\nu \cong 1.1(2)$  obtained is consistent with previous estimates of $\nu$ for the entanglement transition in monitored circuits with Haar-random gates~\cite{zabalo_critical_2020}. With this estimate of $p_m^{c}$, we verify~\cite{supp_mat} that the dynamics at $p_m^{c}$ are consistent with the exponent $z=1$ by examining the purification of an initial mixed state~\cite{gullans_scalable_2020, gullans_dynamical_2020,Iadecola_dynamical}. Thus, along the line $\pf = \pme$, the entanglement and absorbing-state transitions are well separated, and the former displays the same critical properties as monitored circuits without feedback.

\begin{figure}
\centering
\includegraphics[width=\columnwidth]{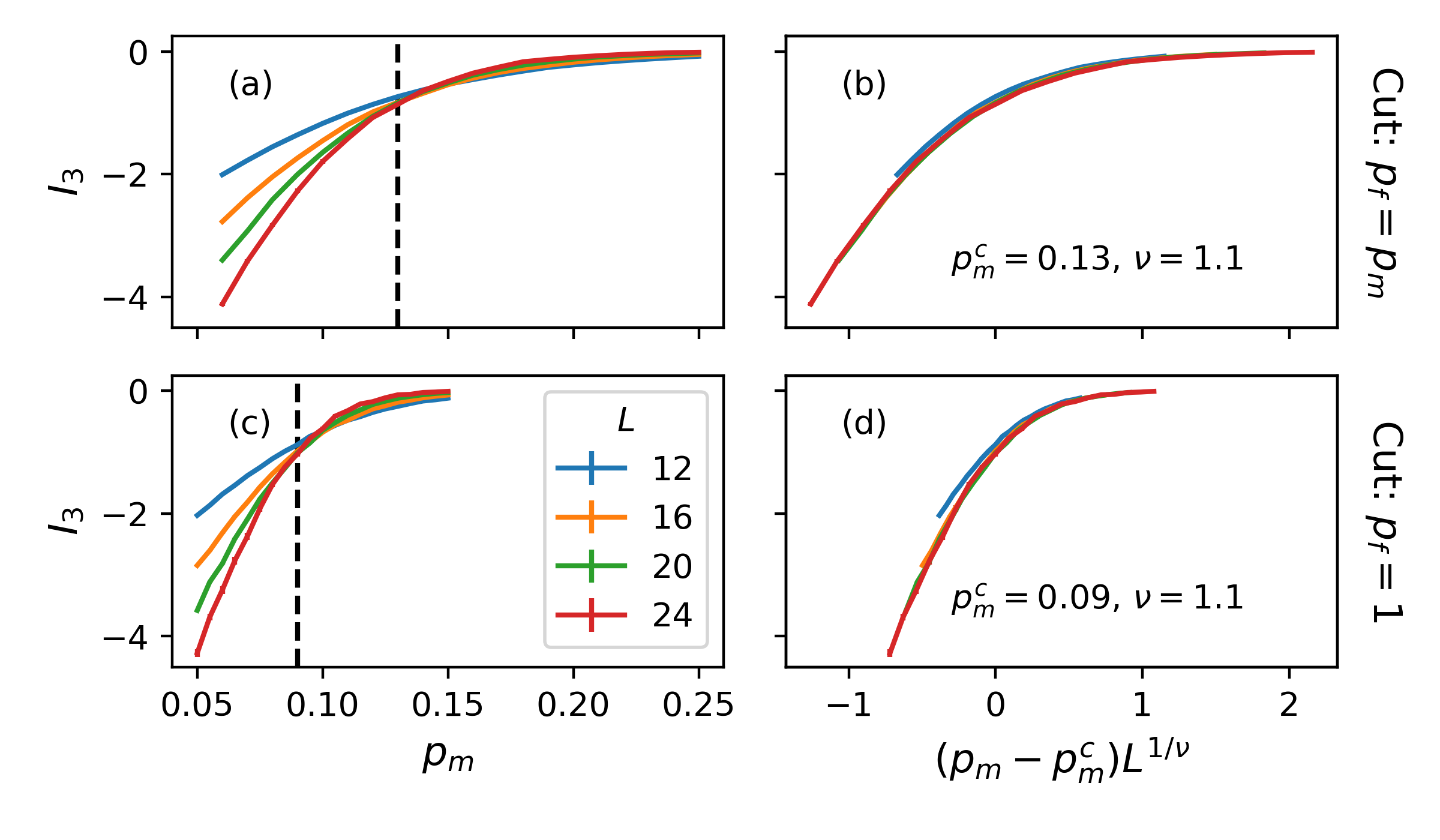}
\caption{
Tripartite quantum mutual information. The data are averaged over $5\cdot 10^2$ - $5\cdot 10^3$ circuit realizations and quantum trajectories in simulations of full-wavefunction evolution, and over times $t=2L$ and $2L+\frac{1}{2}$ to avoid even-odd effects. The top row corresponds to the cut $\pf = \pme$, and the bottom to $\pf=1$. Vertical dashed lines mark estimates of the critical points: $\pme^c \cong 0.130(5)$ (a) and $\pme^c \cong 0.090(5)$ (c). Panels (b) and (d) show the finite-size scaling collapse of the form $I_3 = f[(\pme - \pme^c) L^{1/\nu}]$, with $\nu\cong 1.1(2)$.
\label{fig:I3_allup}}
\end{figure}

\begin{figure}
\centering
\includegraphics[width=\columnwidth]{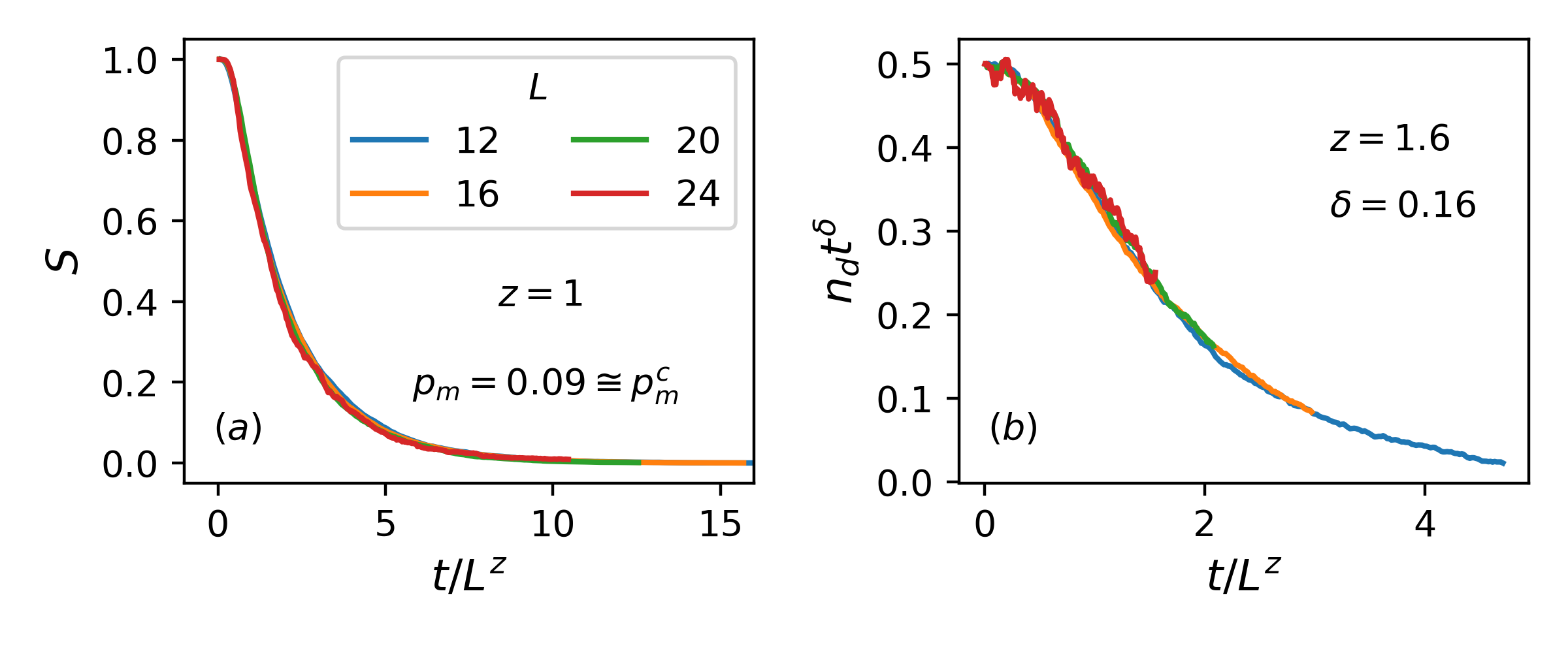}
\caption{Dynamical critical scaling along $\pf=1$. The system is initialized in a mixed state with 1 bit of entropy as described in the SM. Data points are averages of $2\cdot 10^2$ - $5\cdot10^3$ samples. (a) The second Rényi entropy of the density matrix $S$ vs. scaled time. Data collapse is consistent with the scaling form $S = f(t/L)$. (b) The scaled density of defects vs. scaled time. Data collapse is consistent with $n_d = t^{-\delta} \Phi(0, t/L^z)$ at the critical point [Eq.~\ref{eq:n_d_scaling}]. The $z$ values are not consistent in (a) and (b), indicating distinct critical phenomena.
\label{fig:critical_dynamics}}
\end{figure}

We now consider the case of full-strength feedback and vary $\pme$ while $\pf=1$ is fixed. In this case, every measure-and-feedback operation leaves the spin in the $\ket{\uparrow}$ state; hence measurements not only drive trajectories to low-entanglement states, but also specifically towards the polarized state. As a result, the entanglement and absorbing-state transitions come closer together and a similar analysis of $I_3$ locates the entanglement transition at $\pme^c \cong 0.090(5)$ [see Fig.~\ref{fig:I3_allup}(c)], numerically indistinguishable from the location of the absorbing-state transition found above (Fig.~\ref{fig:nd_classical}). 
A natural question is then whether the two phase transitions coalesce as $\pf\rightarrow 1$, or remain distinct critical phenomena. 
In order to see that the two transitions remain distinct, we determine their dynamical exponents $z$~\footnote{While the $I_3$ data along $\pf=1$ still collapses well with $\nu\cong 1.1(2)$ [Fig.~\ref{fig:I3_allup}(d)], this value of $\nu$ is also consistent with the value $\nu_\perp = 1.1$ for directed percolation and thus the absorbing-state transition.}. We extract $z$ for the entanglement transition using the purification setup~\cite{supp_mat}---which examines the entropy $S$ of a system initially in a mixture of two random orthogonal states---and for the absorbing-state transition using the density of defects $n_d$.
In Fig.~\ref{fig:critical_dynamics} we show both $S$ and $n_d$ data for times $t \gtrsim L^z$. The entropy continues to scale like $S\sim f(t/L)$, i.e., $z=1.0(1)$, and $n_d$ scales according to $\delta = 0.16(8)$ and $z=1.6(1)$ corresponding to directed percolation. The picture that emerges is one in which the two phase transitions can become close in $\pme$ at strong feedback ($\pf = 1$); however, they remain distinct critical phenomena that happen on parametrically different time-scales as indicated by the different $z$ values.

\emph{\textbf{Argument for Separate Transitions---}}
We now argue more directly that a sliver of Phase II generically separates the two transitions, provided that the absorbing-state transition is continuous.
We perform a Stinespring dilation~\cite{nielsen2002quantum} of the channel at $\pf=1$, such that each measurement consists of swapping out one of the system spins for an ancilla spin prepared in $\ket{\uparrow}$. The steady-state entropy of $\rhoavg$ is the entanglement between the system and ancilla qubits; meanwhile, the entanglement entropy of trajectories is that of the system's wavefunction once we measure all the ancillas.
Consider the dynamics when $n_d \ll 1$. ``Live'' (spin-$\downarrow$) regions are dilute, and collisions between them are rare. However, processes where a qubit in the live region is swapped into the environment are common, since $\pme$ is $O(1)$. Suppose the live region is initially entangled with the rest of the system. Before it encounters a neighbor, it undergoes many measurement/feedback processes. Therefore, the rest of the system is now entangled with the live region together with the qubits that were swapped out. Collisions are rare, so the number of swapped-out qubits exceeds the number of qubits that remain in the live region. By the decoupling principle~\cite{schumacher2002approximate}, the live region is decoupled from the rest of the system: instead, the rest of the system is now entangled with the environment. Since decoupling happens between collisions, large-scale entanglement cannot build up and the system remains in an area-law phase. Crucially, the rate of collisions (set by $n_d$) can be made arbitrarily small relative to the rate of swaps (set by $\pme$), by going near the absorbing-state transition~\footnote{In our numerics, $n_d$ at the critical point is still relatively high at the accessible system sizes, so this asymptotic separation might not be visible.}.

\emph{\textbf{SPT Absorbing State---}} 
In principle, the approach described above can be generalized to steer a system to any pure state that is the unique state annihilated by a set of projectors. 
However, the efficacy of such steering depends on the nature of correlations in the target state: 
as noted in the introduction, local quantum channels have light cones and cannot prepare states with long-range correlations in finite time starting from a product state. 
Feedback operations targeting states with long-range correlations are inherently less efficient, since they generically move defects instead of eliminating them. One might expect, therefore, that the entanglement and absorbing-state transitions should be well-separated for such target states even at $\pf = 1$. 

We have tested this intuition for dynamics targeting the cluster state (ground state of $H = - \sum_i Z_i X_{i+1} Z_{i+2}$), which is an SPT state; under unitary dynamics, this state can be prepared in $O(1)$ depth if the protecting $Z_2\times Z_2$ symmetry (that is a product of $X_i$ on even/odd sites) is broken, while it requires depth $O(L)$ in the presence of the symmetry. An appealing feature of the cluster state is that it allows steering protocols of two types: (a)~those in which the feedback completely breaks the protecting symmetry, allowing individual defects to be annihilated, and (b)~those in which the feedback respects a parity symmetry that constrains defects to be annihilated in pairs. The details of the interactive dynamics are described in~\cite{supp_mat}; the results can be summarized as follows. In case~(a), the phase diagram and dynamics closely resemble the product-state example, while case~(b) has the following salient differences. First, the entanglement and absorbing-state transitions are widely separated even for $\pf = 1$ [Fig.~\ref{fig:circuit_and_phasediagram}(d)]. Second, the absorbing-state transition belongs to the ``parity conserving" universality class~\cite{supp_mat}. Even in the absorbing phase, the approach to steady state takes a time that scales as $L^2$ because of the diffusion-limited recombination of defects. This conclusion does not rely on SPT symmetry, and we expect it to generalize to other models involving pairwise annihilation of defects, including in higher dimensional systems with more exotic anyonic defects. Third, in the area-law non-absorbing regime,  individual trajectories possess ``spin-glass'' SPT order~\cite{lavasani_measurement-induced_2021, ippoliti_entanglement_2021, Sang_2021}, but the average over measurement outcomes washes out this order. Finally, with the parity symmetry, if we allow nonlocal communication so defects can be paired up more efficiently, directed-percolation universality is restored.

\emph{\textbf{Discussion---}} In this work we explored the dynamics of single trajectories and the trajectory-averaged density matrix for a family of interactive quantum circuits with measurements and feedback. We found distinct phase transitions for single-trajectory and trajectory-averaged quantities, and argued that in general these transitions should belong to distinct universality classes and occur at distinct critical measurement rates. The feedback allowed in our setup was based on \emph{local} information about the state; thus it was much more restrictive than the types of feedback that are allowed in general LOCC protocols or in quantum error correction. An interesting direction for future work is to explore the consequences of relaxing this locality constraint, and allowing for general forms of interactive dynamics that cannot be captured by local quantum channels (and therefore obey weaker locality constraints). It would also be interesting to understand cases where the two transitions may coincide~\cite{Buchhold_preselection}, and the relevance of feedback to these transitions in an RG sense.

\emph{\textbf{Note Added---}} In the final stages of completion of this work, we became aware of related work which appeared on the arXiv and was subsequently published as~\cite{ChenAdaptive2}. Our results agree where they overlap. 

\emph{\textbf{Acknowledgements---}} We are grateful to Tibor Rakovszky for many insightful discussions and close collaboration. We also thank Sebastian Diehl for discussing his work with us, and Yaodong Li and David Huse for helpful discussions. This work was supported by the US Department of Energy, Office of Science, Basic Energy Sciences, under Early Career Award Nos. DE-SC0021111 (N.O.D and V.K.).  V.K. also acknowledges support from the Alfred P. Sloan Foundation through a Sloan Research Fellowship and the Packard Foundation through a Packard Fellowship in Science and Engineering. S.G. acknowledges support from NSF DMR-1653271. A.M. is supported in part by the Stanford Q-FARM Bloch Postdoctoral Fellowship in Quantum Science and Engineering and the Gordon and Betty Moore Foundation’s EPiQS Initiative through Grant GBMF8686.  Numerical simulations were performed on Stanford Research Computing Center's Sherlock cluster.  We acknowledge the hospitality of the Kavli Institute for Theoretical Physics at the University of California, Santa Barbara (supported by NSF Grant PHY-1748958). 

\bibliography{main}

\end{document}

\title{Supplemental Material for ``Entanglement and Absorbing-State Transitions in Interactive Quantum Dynamics"}

\author{Nicholas O'Dea}
\affiliation{Department of Physics, Stanford University, Stanford, CA 94305, USA}

\author{Alan Morningstar}
\affiliation{Department of Physics, Stanford University, Stanford, CA 94305, USA}

\author{Sarang Gopalakrishnan}
\affiliation{Department of Electrical Engineering, Princeton University, Princeton, NJ 08544, USA}

\author{Vedika Khemani}
\affiliation{Department of Physics, Stanford University, Stanford, CA 94305, USA}

\date{\today}

% \begin{abstract}
% An abstract for the supp mat, if needed.
% \end{abstract}

\maketitle

\section{Classical stochastic process for the dynamics of $\rhoavg$ for the polarized absorbing state}
\label{app:classical_polarized}

In this section, we describe the mapping of the dynamics of the diagonal of the averaged density matrix to a classical stochastic process for the absorption to the polarized absorbing state. The mapping follows from consideration of the channel describing the time evolution of the averaged density matrix. This channel can be described as a circuit acting on the averaged density matrix in a ``doubled" Hilbert space, with rows of single-site operators corresponding to $Z$ measurements and rows of two-site operators corresponding to the constrained Haar unitary evolution which locally preserves $\ket{\uparrow\uparrow}$.

The form of the single-site averaged measurement operator $O_m$ follows from the evolution of the averaged density matrix at a given site (site index $i$ suppressed) during the measurement round:
\begin{align}
    \rhoavg &\to (1-p_{\rm m}) \rhoavg + p_{\rm m} \left(P_\uparrow \rhoavg P_\uparrow  + (1-p_{\rm f})P_\downarrow \rhoavg P_\downarrow + p_{\rm f} X P_\downarrow \rhoavg P_\downarrow X \right)
\end{align}
where $P_{\uparrow}$ ($P_{\downarrow}$) is a projector onto spin up (down). In the doubled Hilbert space,
\begin{equation}
    O_{m} = \begin{pmatrix}
    1 & 0 & 0 & p_{\rm m}p_{\rm f} \\
    0 & 1-p_{\rm m} & 0 & 0 \\
    0 & 0 & 1-p_{\rm m} & 0 \\
    0 & 0 & 0 & 1-p_{\rm m}p_{\rm f} \\
    \end{pmatrix}
\end{equation}
Here we use the usual basis of $\{\ket{\uparrow}\ket{\uparrow }, \ket{\uparrow} \ket{\downarrow }, \ket{\downarrow} \ket{\uparrow }, \ket{\downarrow} \ket{\downarrow}\}.$ Note that this operator does not mix off-diagonal terms of the density matrix with diagonal ones; it is block-diagonal with the blocks 
\begin{equation}
\begin{split}
\label{eq:pol_meas_blockdiagonal}
    O_{m}^{(d)} &=  \begin{pmatrix}
    1 & p_{\rm m}p_{\rm f} \\
    0 & 1-p_{\rm m}p_{\rm f}
    \end{pmatrix}\\
    O_{m}^{(o)} &=  \begin{pmatrix}
    1-p_{\rm m} & 0 \\
    0 & 1-p_{\rm m}
    \end{pmatrix}    
\end{split}
\end{equation}
Here the basis vectors are taken as $\{\ket{\uparrow} \ket{\uparrow }, \ket{\downarrow}\ket{\downarrow }\}$ and  $\{\ket{\uparrow } \ket{\downarrow}, \ket{\downarrow} \ket{\uparrow}\}$ respectively.

The average of the constrained Haar unitary is similarly block-diagonal. We have 
\begin{equation}
\label{eq:pol_unit_blockdiagonal}
\begin{split}
    O_{U}^{(d)} &=  \begin{pmatrix}
    1 & 0 & 0 & 0 \\
    0 & \frac{1}{3} & \frac{1}{3} & \frac{1}{3} \\
    0 & \frac{1}{3} & \frac{1}{3} & \frac{1}{3} \\
    0 & \frac{1}{3} & \frac{1}{3} & \frac{1}{3}
    \end{pmatrix}  
\end{split}
\end{equation}
in the basis $\{\ket{\uparrow \uparrow} \ket{\uparrow \uparrow},  \ket{\uparrow \downarrow } \ket{\uparrow \downarrow }, \ket{\downarrow \uparrow} \ket{\downarrow \uparrow}, \ket{\downarrow \downarrow}\ket{\downarrow \downarrow}\}$, while the block describing the off-diagonal elements of the averaged density matrix vanishes.

From the explicit forms of $O_{m}$ and $O_{U}$, we see that the channel does not mix the diagonal and off-diagonal elements of the averaged density matrix $\rho$, and so the diagonal elements only transform among themselves. We note that after a single layer of unitaries, all the off-diagonal elements are set to zero by the vanishing of $O_{U}^{(o)}$. This vanishing is not strictly needed for the validity of this classical mapping, as to probe the absorbing state transition via diagonal operators like the magnetization, one only needs to consider the diagonal elements of the averaged density matrix. 

We see from Eqs.~\ref{eq:pol_meas_blockdiagonal} and ~\ref{eq:pol_unit_blockdiagonal} that the time evolution of the diagonal of the reduced density matrix is equivalent to the transfer matrix of a classical stochastic process in the style of a cellular automaton. Specifically, a diagonal element of the density matrix is identified with a classical bitstring (e.g. $\ket{\uparrow \uparrow \downarrow \uparrow}\bra{\uparrow \uparrow \downarrow \uparrow}$ is identified with $0010$). In a measurement round, if a site of the bitstring is $1$, we see from Eq.~\ref{eq:pol_meas_blockdiagonal} that the site stays the same with probability $1-p_{\rm m}p_{\rm f}$ or flips to $0$ with probability $p_{\rm m}p_{\rm f}$. Similarly, in the unitary layers, we see from Eq.~\ref{eq:pol_unit_blockdiagonal} that $00$ is preserved while $11$, $10$, $01$ mix among each other with equal probability $1/3$. As noted above, this exact mapping allows us to calculate quantum expectation values of diagonal operators by averaging over many realizations of this classical stochastic process. Furthermore, we see that the number of parameters describing the dynamics is reduced on considering the averaged density matrix; $p_{\rm m}$ and $p_{\rm f}$ only enter in the product $p_{\rm m}p_{\rm f}$.

This classical mapping can be straightforwardly generalized. We note that quantum steps in the original Hilbert space involving the action of Pauli strings are particularly simple to convert to the classical dynamics. This is because a diagonal element of the density matrix transforms to another diagonal element under the action of conjugation by a Pauli string - $X$ and $Y$ flip spins, while the phases in $Y$ and $Z$ cancel between the two copies of the state. This means that $X$ and $Y$ terms in the Pauli string correspond to flipping a site of the classical bitstring, while $Z$ and $I$ do nothing. For example, the diagonal element $\ket{\uparrow \uparrow \downarrow \uparrow}\bra{\uparrow \uparrow \downarrow \uparrow}$ transformed by $X_1I_2 Z_3I_4$ is $X_1I_2 Z_3I_4 \ket{\uparrow \uparrow \downarrow \uparrow}\bra{\uparrow \uparrow \downarrow \uparrow}X_1I_2 Z_3I_4 = (-1)^2 \ket{\downarrow \uparrow \downarrow \uparrow}\bra{\downarrow \uparrow \downarrow \uparrow}$. This corresponds to the transformation of the classical bitstring $0010$ to $1010$. We use this freely in our discussion of classical stochastic models for our dynamics that absorb into the cluster state (see below).

\section{Purification dynamics}

\begin{figure}
\centering
\includegraphics[width=0.5\columnwidth]{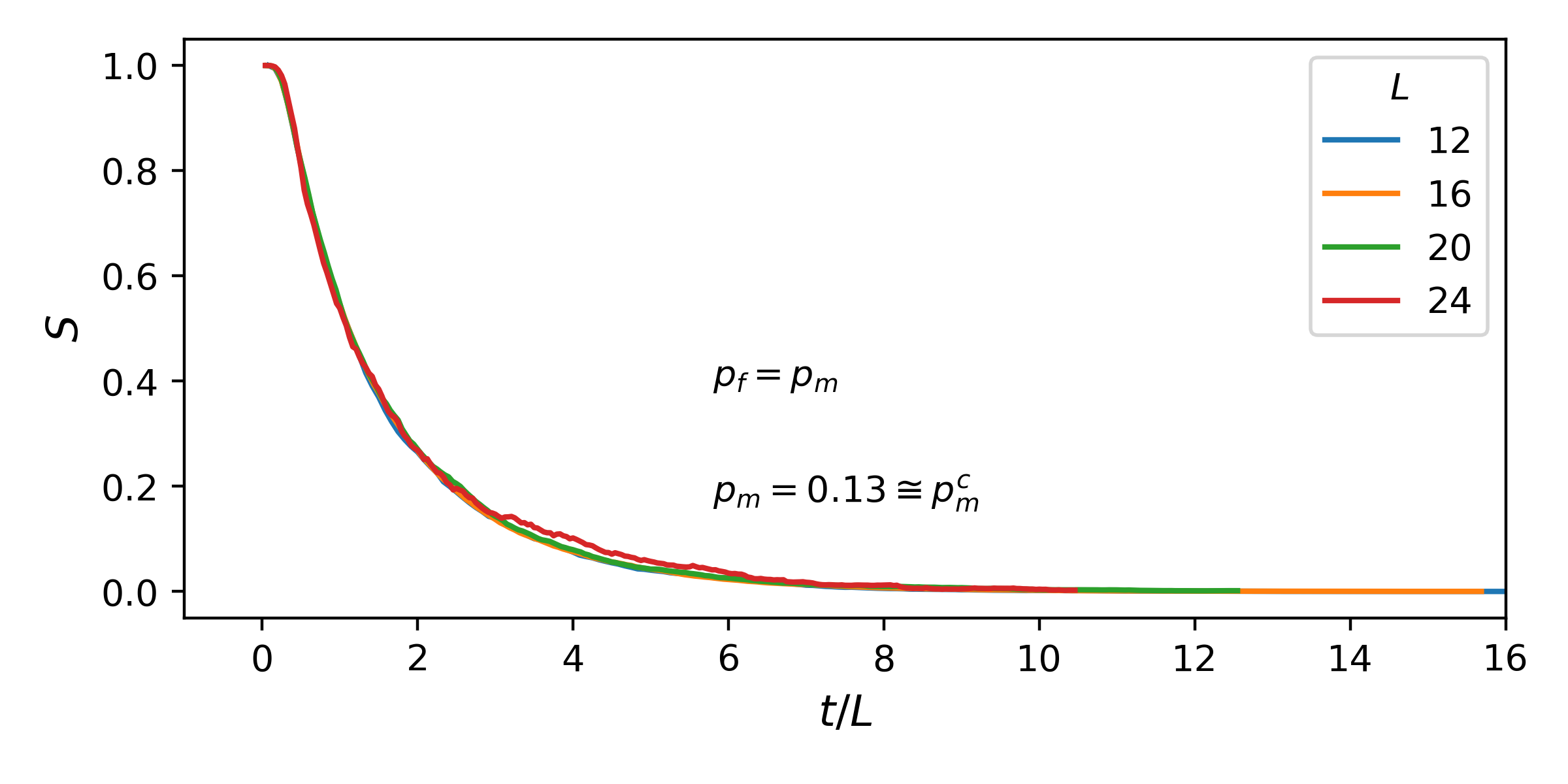}
\caption{Dynamical critical scaling along $\pf=\pme$. The system is initialized in a mixed state with 1 bit of entropy as described in the SM. Data points are averages of $2\cdot 10^2$ - $5\cdot10^3$ samples. Shown is the second Rényi entropy of the density matrix $S$ vs. scaled time. Data collapse is consistent with the scaling form $S = f(t/L)$, i.e., $z=1$ for the entanglement transition that is well separated from the absorbing-state transition, as expected.}
\label{fig:S_scaling}
\end{figure}

Here we lay out our setup for using the dynamics of purification to estimate the dynamical critical exponent $z$ in Fig.~\ref{fig:S_scaling} and in the main text. 
We also use this setup to estimate the location of the phase transitions in our models that target the cluster state in Fig.~\ref{fig:ZXZ_models} and sections below.

We initialize our system in a mixed state of two random orthogonal pure states 
\begin{align}
    \rho = \frac{1}{2}(|\psi\rangle\langle\psi| + |\psi_\perp\rangle\langle\psi_\perp|)
    \label{eq:mixed_state}
\end{align}
and track the second Rényi entropy of the density matrix $S=-\log_2 \Tr{\rho^2}$ in time, averaging over samples. This is similar in spirit to the strategy of starting in an initial fully mixed state $\rho \propto I$~\cite{gullans_dynamical_2020}, however in this setup the density matrix begins with 1 bit of entropy, rather than $L$ bits. Our setup is also approximately equivalent to initializing a spin of the (pure) system in a Bell pair with an ancilla spin, evolving the system for an extensive time under purely unitary dynamics, then evolving under the monitored dynamics while tracking the Rényi entanglement entropy of the ancilla~\cite{gullans_scalable_2020}. Since $S=1$ initially, rather than $S=L$, as is the case when the initial density matrix has full rank, we expect the scaling form $S \sim f(t/L^z)$, with $z=1$, where $f(\cdot)$ decays faster than a power law~\cite{gullans_dynamical_2020}. 

In the main text we employ this setup to extract the critical exponent $z$ by fixing $\pf$ and $\pme$ to a critical pair of values and collapsing the data $S$ vs. $t/L^z$ for various $L$ using the correct choice of $z$. For the case of $\pme = \pf \cong 0.13$, we indeed find a good collapse with $z=1.0(1)$ shown in Fig.~\ref{fig:S_scaling}. This known result is not surprising because the entanglement transition is well separated from the absorbing-state transition along the cut $\pf = \pme$. We also use the same approach in the main text to show that $z=1$ remains true at the critical point along $\pf=1$, where the entanglement and absorbing-state phase transitions are at numerically indistinguishable locations.

\section{Targeting the cluster state \label{sec:target_cluster}}

One can achieve a cleaner separation between the entanglement and absorbing-state transitions by considering systems in which the approach to the absorbing state is frustrated by the presence of nontrivial correlations.
To this end, we consider feedback that drives the system into the cluster state, an entangled SPT state which is the unique +1 eigenstate of the set of $\{ \mathcal{C}_i\equiv Z_{i-1} X_i Z_{i+1} \}$ stabilizers. The cluster state can be prepared with an $O(1)$ depth unitary circuit if the protecting $Z_2\times Z_2$ symmetry (that is a product of $X_i$ on even/odd sites) is broken, while it requires depth $O(L)$ in the presence of the symmetry. 
Instead, we again consider interactive dynamics with measurements of $\mathcal{C}_i$ that occur with probability $p_{\rm m}$, and feedback with probability $p_{\rm f}$ (see Fig.~\ref{fig:cluster_circuit} and Sec.~\ref{app:cluster_model} for circuit architecture and model details). We will use scrambling three-site unitary gates $U_{i-1, i, i+1}$  that preserve the cluster state and additionally commute with the symmetry $\prod_i Z_{i-1} X_{i}Z_{i+1} = \prod_i X_i$ which ensures that the unitaries conserve the parity of the number of incorrect stabilizers. While this symmetry is weaker than the full $Z_2\times Z_2$ symmetry, the scrambling unitary circuit is not designed to efficiently reach the cluster state absent measurements. In this and the next sections, we demonstrate that certain choices of feedback can yield widely separated entanglement and absorbing state transitions, even at $\pf=1$, and can change the universality class of the absorbing state transition.

\begin{figure}
\centering
\includegraphics[width=0.6\columnwidth]{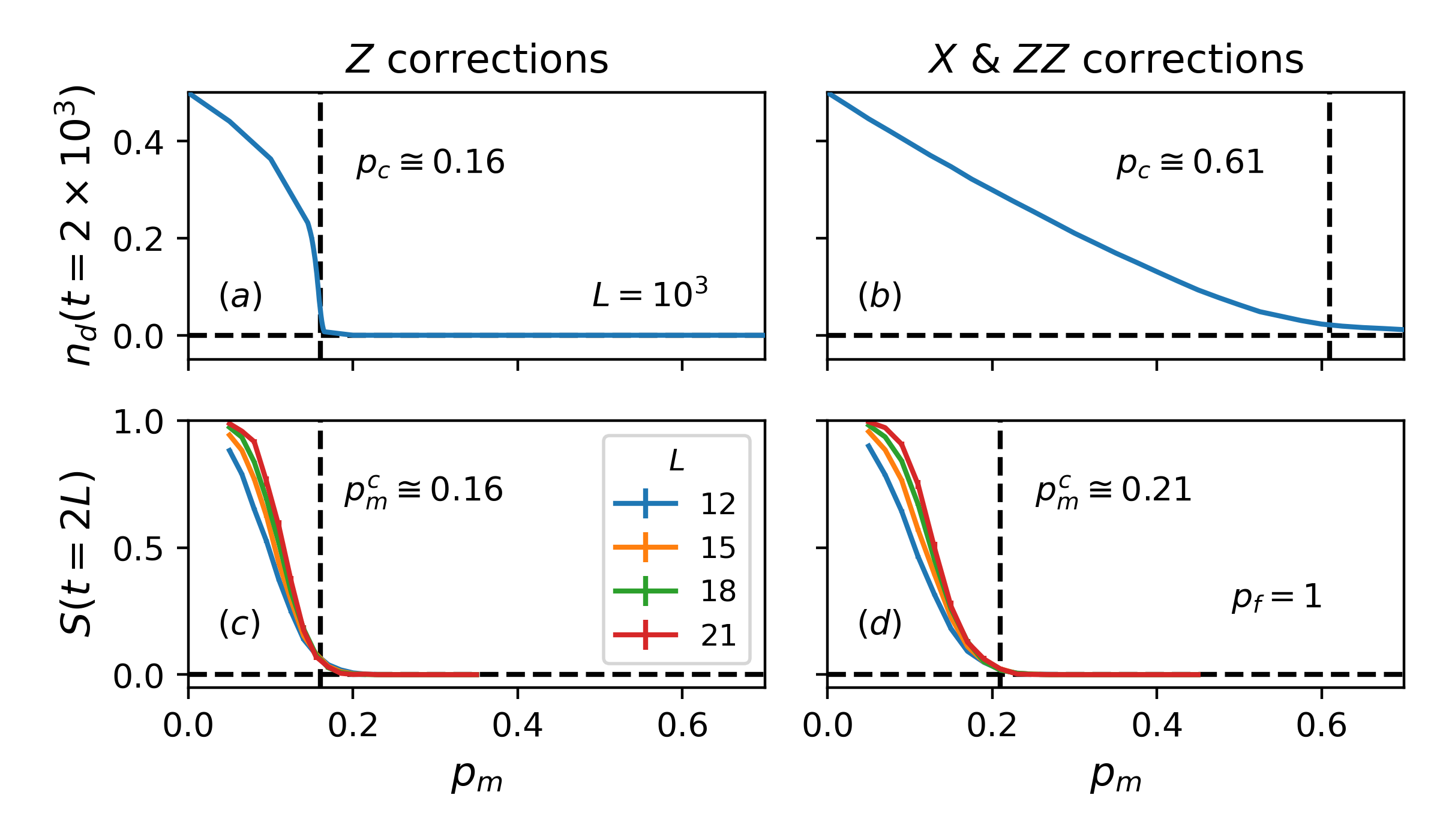}
\caption{Absorbing-state and entanglement transitions in circuits that target the cluster state along $\pf=1$. The left (right) column corresponds to the model with $Z$ ($ X\ \&\ ZZ$) corrective operations. The top row contains $n_d$ data obtained via the mapping to a classical stochastic model. The bottom row contains $S$ data at $t=2L$ obtained by simulating the evolution of an initially mixed density matrix with one bit of entropy. The vertical dashed lines indicate estimates of the critical measurement rates for the absorbing-state (a,b) and  entanglement (c,d) transitions. In the case of $X\ \&\ ZZ$ corrections, the transitions are evidently separate.
\label{fig:ZXZ_models}}
\end{figure}

First, consider the case where the feedback \emph{breaks} the parity and SPT symmetries, and corrects a measurement of $\mathcal{C}_i = -1$ by acting with $Z_i$ (which flips $\mathcal{C}_i$). In this case, at $\pme = \pf=1$, the cluster state is prepared in one time step. The phase diagram is qualitatively similar to that discussed above for the polarized state. In particular, the two transitions are again numerically coincident at $\pf = 1$ [Fig.~\ref{fig:ZXZ_models}(a,c)], but differ in their critical data. 

Second, consider symmetry-preserving feedback in which a measurement of $\mathcal{C}_i = -1$ is followed by randomly acting with either $X_{i-1}$ (flips both $\mathcal{C}_i$ \emph{and} $\mathcal{C}_{i-2})$ or $X_{i+1}$ (flips both $\mathcal{C}_i$ \emph{and} $\mathcal{C}_{i+2})$. Thus, feedback either hops defects or only annihilates them in \emph{pairs} on the even/odd sublattices. In this case, even at $\pme = \pf = 1$, the cluster state cannot be prepared in any finite time since pairing up defects over long distances requires non-local classical communication which is not a resource available to the local quantum channel. This gives a knob --- which exploits long-range correlations and locality --- to separate the entanglement and absorbing-state transitions: while frequent measurements prepare area-law states, the feedback is less effective at simultaneously steering these towards the target state. 
It turns out that this model works `too well' at separating the transitions, as we numerically find that the system is not in the absorbing phase even at $\pme=\pf =1$ (however, different architectures with more measurement layers between unitaries may be made absorbing). To mitigate this, we consider a different feedback model which also allows for $Z_{i-1}Z_i$ or $Z_{i} Z_{i+1}$ as feedback unitaries (together with the $X_{i-1/i+1}$'s, with each of the four feedback options acting with equal probability). This allows neighboring defects to also get annihilated in pairs, thereby breaking the $Z_2\times Z_2$ symmetry but preserving the overall parity (like the unitary part of the evolution). In this case, we do get an absorbing-state transition, but it belongs to the  ``parity conserving" universality class. In this case, $n_d$ approaches zero in the absorbing phase only diffusively in time, in contrast to the exponential decay for the DP class. 

For each of the two models above, we estimate the location of both the absorbing-state and entanglement phase transitions along the cut $\pf=1$. To do this we again employ a mapping to a classical stochastic model to study the absorbing-state dynamics [Sec.~\ref{app:classical_cluster}], and exact simulations of the full wavefunction to locate the entanglement transition. Fig.~\ref{fig:ZXZ_models} shows the entanglement and absorbing-state transitions~\footnote{For the entanglement transition, we do not locate the critical point using the tripartite mutual information $I_3$ as was done in the main text because our cluster-state models have a 3-site brickwork circuit architecture and $I_3$ requires $L$ to be a multiple of 4; this is an inconvenient pair of constraints given our small range of $L$. Instead, we initialize our system in an even mixture of two random orthogonal pure states and compute the Rényi entropy $S$ of the density matrix at time $t=2L$.}. The main result is the wide separation of the two transitions in the case of $X$ and $ZZ$ corrections, where the symmetry $\prod_i X_i$ is preserved, in contrast to the case with $Z$ corrective operations where the two transitions are again numerically coincident. This more general analysis does not rely on SPT symmetry, and we expect it to bear out in other parity-conserving models involving pairwise annihilation of defects, including in higher dimensional systems with more exotic anyonic defects. 

\section{Models for the cluster state dynamics}
\label{app:cluster_model}

\begin{figure}
    \centering
    \includegraphics[width=0.33\columnwidth]{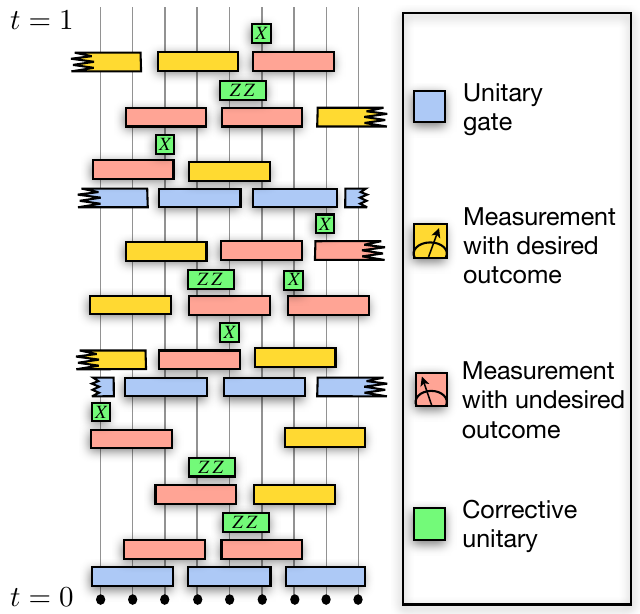}
    \caption{Adaptive circuit targeting the cluster ground state of $H=-\sum_i Z_{i-1}X_{i}Z_{i+1}$. One period of time evolution is shown. The target state is stabilized by all $Z_{i-1}X_{i}Z_{i+1}$. Measurements of the stabilizers are followed by corrective unitary operations. The random unitary gates are described in the main text and respect the symmetry $\prod_i X_i$.}
    \label{fig:cluster_circuit}
\end{figure}

We discuss here the explicit form of our cluster state dynamics. We first fix our architecture as a brickwork circuit with three-site gates, as in Fig.~\ref{fig:cluster_circuit}.
One unit of time comprises three layers of unitaries, each shifted by one site, with measurements following each unitary layer.  The first unitary layer comprises random three-site unitary gates (described below) acting on sites $(3i, 3i+1, 3i+2)$, the next has unitaries on sites $(3i+1, 3i+2, 3i+3)$ and the last on sites 
$(3i+2, 3i+3, 3i+4)$, with $i=0,\cdots L$ and periodic boundary conditions. Each unitary layer is followed by a three-layer brickwork of three-site $Z_{i-1} X_i Z_{i+1}$ measurement and feedback operations, with the measurement layers again displaced relative to each other. 
Thus, each stabilizer has the opportunity to be measured with probability $\pme$ following each unitary layer. Given an outcome of $-1$, the measurement is immediately followed by feedback occurring with probability $p_{\rm f}$ and supported on the measurement locations, with the choice of feedback varying between the models.

For all of our models with absorption into the cluster state, we consider three-site unitaries $U_{i-1, i, i+1}$ centered on $i$ that preserve the cluster state and $\prod_i Z_{i-1} X_{i}Z_{i+1} = \prod_i X_i$. The latter symmetry preserves the parity of the defects. A general, random unitary satisfying this form is 
\begin{equation}
U = 
\begin{pmatrix}
I  &
\begin{matrix}
0 & 0 & 0 & 0 \\
0 & 0 & 0 & 0 \\
0 & 0 & 0 & 0 \\
0 & 0 & 0 & 0
\end{matrix} \\
\begin{matrix}
0 & 0 & 0 & 0 \\
0 & 0 & 0 & 0 \\
0 & 0 & 0 & 0 \\
0 & 0 & 0 & 0
\end{matrix} &
\begin{matrix} U(2) & 
\begin{matrix} 0 & 0 \\ 0 & 0 \end{matrix} \\
\begin{matrix} 0 & 0 \\ 0 & 0 \end{matrix} &  U(2)
\end{matrix}
\end{pmatrix}
\label{eq:ZXZ_unitary}
\end{equation}
Note that the above matrix is written in a transformed basis relative to the computational basis that we use in our simulations. To write the matrix above in the computational basis, take $U_{\rm comp} = V^\dagger U V$ with 
\begin{equation}\label{eq:ZXZ_unitary_basis}
V = 
\begin{pmatrix}
 \frac{1}{\sqrt{2}} & 0 & \frac{1}{\sqrt{2}} & 0 & 0 & 0 & 0 & 0 \\
 0 & \frac{1}{\sqrt{2}} & 0 & -\frac{1}{\sqrt{2}} & 0 & 0 & 0 & 0 \\
 0 & 0 & 0 & 0 & 0 & \frac{1}{\sqrt{2}} & 0 & \frac{1}{\sqrt{2}} \\
 0 & 0 & 0 & 0 & \frac{1}{\sqrt{2}} & 0 & -\frac{1}{\sqrt{2}} & 0 \\
 0 & \frac{1}{2} & 0 & \frac{1}{2} & -\frac{1}{2} & 0 & -\frac{1}{2} & 0 \\
 \frac{1}{2} & 0 & -\frac{1}{2} & 0 & 0 & \frac{1}{2} & 0 & -\frac{1}{2} \\
 0 & \frac{1}{2} & 0 & \frac{1}{2} & \frac{1}{2} & 0 & \frac{1}{2} & 0 \\
 \frac{1}{2} & 0 & -\frac{1}{2} & 0 & 0 & -\frac{1}{2} & 0 & \frac{1}{2} \\

\end{pmatrix}
\end{equation}
The forms of $U$ and $V$ above follow from the constraints of preserving the cluster state and the parity of defects. 

First, to preserve the cluster state, the three-site unitary must act proportionally to the identity on the subspace corresponding to the four eigenvectors of the cluster state's three-site reduced density matrix (i.e. the $+1$ eigenvectors of $Z_{i-1}X_iZ_{i+1}$): $\{\ket{\uparrow + \uparrow},\ket{\uparrow - \downarrow},\ket{\downarrow + \downarrow}, \ket{\downarrow - \uparrow} \}$. Here we use the notation $\ket{+} = \frac{\ket{\uparrow} + \ket{\downarrow}}{\sqrt{2}}$ and $\ket{-} = \frac{\ket{\uparrow} - \ket{\downarrow}}{\sqrt{2}}$ to more compactly represent the states. This constraint corresponds to the $4$ by $4$ identity block in $U$ and the first four rows of $V$. 

The $-1$ eigenvectors of $Z_{i-1}X_iZ_{i+1}$ are allowed to partially mix; the constraint of preserving $\prod_i X_i$ means that of the four $-1$ eigenvectors, only those with the same eigenvalue under $\prod_i X_i$ can mix. The subspace of $-1$ eigenvectors of $Z_{i-1}X_iZ_{i+1}$ with eigenvalue $-1$ under $\prod_i X_i$ is spanned by $\{ \frac{\ket{\uparrow + \downarrow} - \ket{\downarrow + \uparrow}}{\sqrt{2}}, \frac{\ket{\uparrow - \uparrow} + \ket{\downarrow - \downarrow}}{\sqrt{2}} \}$, corresponding to fifth and sixth rows of $V$. We allow unconstrained Haar random dynamics within this subsector as denoted by the first $U(2)$ in $U$. Similarly, the subspace of $-1$ eigenvectors of $Z_{i-1}X_iZ_{i+1}$ with eigenvalue $1$ under $\prod_i X_i$ is spanned by $\{ \frac{\ket{\uparrow + \downarrow} + \ket{\downarrow + \uparrow}}{\sqrt{2}}, \frac{\ket{\uparrow - \uparrow} - \ket{\downarrow - \downarrow}}{\sqrt{2}}\}$, corresponding to the final two rows of $V$. Again, we allow unconstrained dynamics within this sector, as denoted by the lower right $U(2)$ block in $U$.

The three models we consider vary in their choice of feedback. These three choices are, given a measurement outcome of $Z_{i-1} X_i Z_{i+1} = -1$ and given that feedback occurs,
\begin{itemize}
    \item $Z_i$
    \item Equal likelihood of $X_{i-1}$, $X_{i+1}$, $Z_{i-1}Z_i$, or $Z_{i}Z_{i+1}$
    \item Equal likelihood of $X_{i-1}$ or $X_{i+1}$
\end{itemize}
The first choice of feedback breaks parity conservation, $\prod_i X_i$, while the latter two preserve it. The last choice of feedback additionally conserves the product of $X$ on even and odd sites separately.

\section{Classical mapping for the cluster state dynamics}
\label{app:classical_cluster}
In this section, we consider the two cluster state models and mappings from the diagonals of their averaged density matrices to classical stochastic processes. Throughout this section, we consider a transformed basis. The basis transformation $U$ corresponds to acting $CZ$ gates on every adjacent pair of sites and then acting Hadamard gates on every site. This choice of $U$ maps the $ZXZ$ stabilizers to $Z$ stabilizers and the cluster state to the fully polarized state. We find that the diagonal and off-diagonal elements of our averaged density matrix decouple in this basis, making it appropriate for our classical mapping procedure.

\begin{figure}
\centering
\includegraphics[width=0.6\columnwidth]{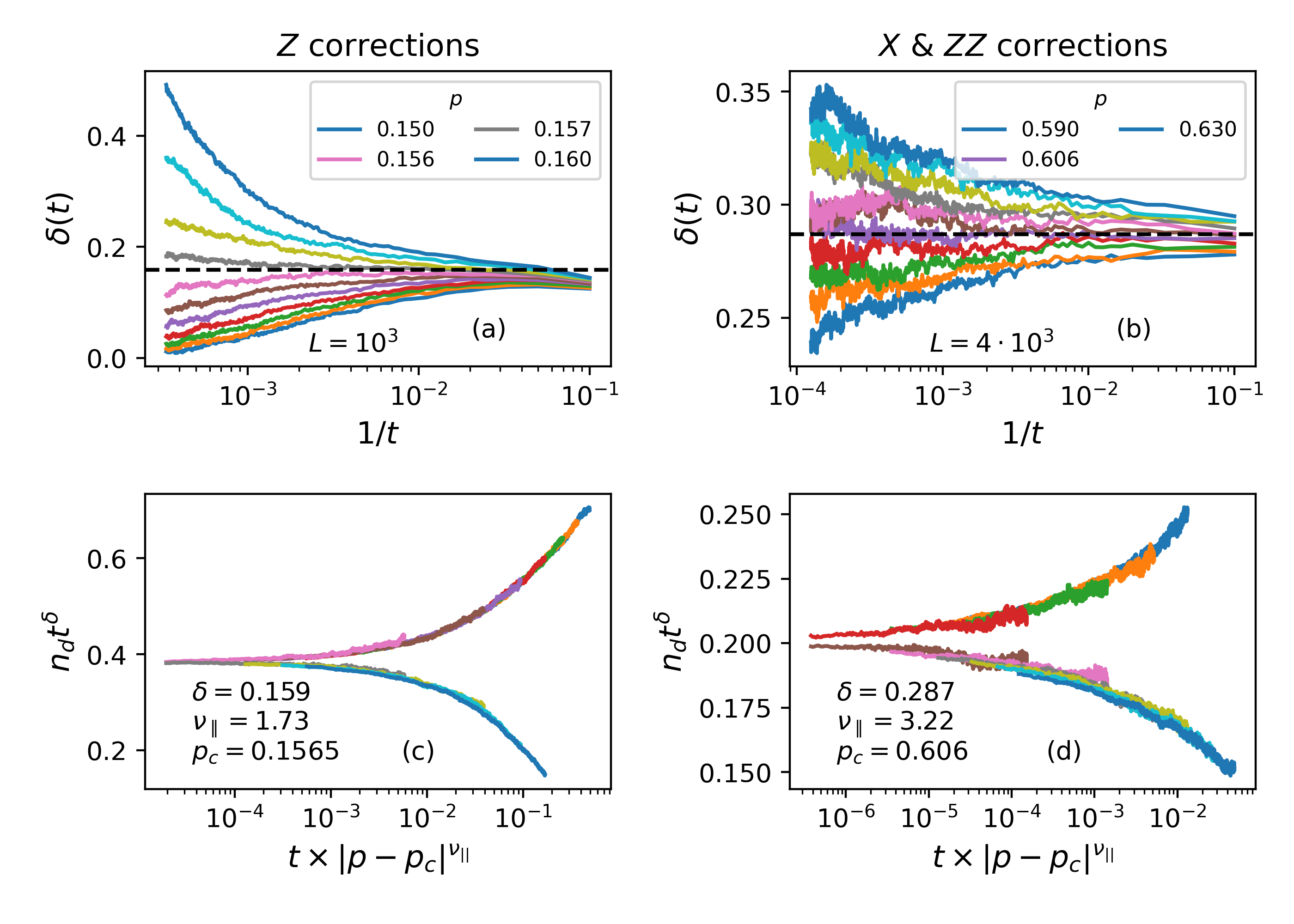}
\includegraphics[width=0.33\columnwidth]{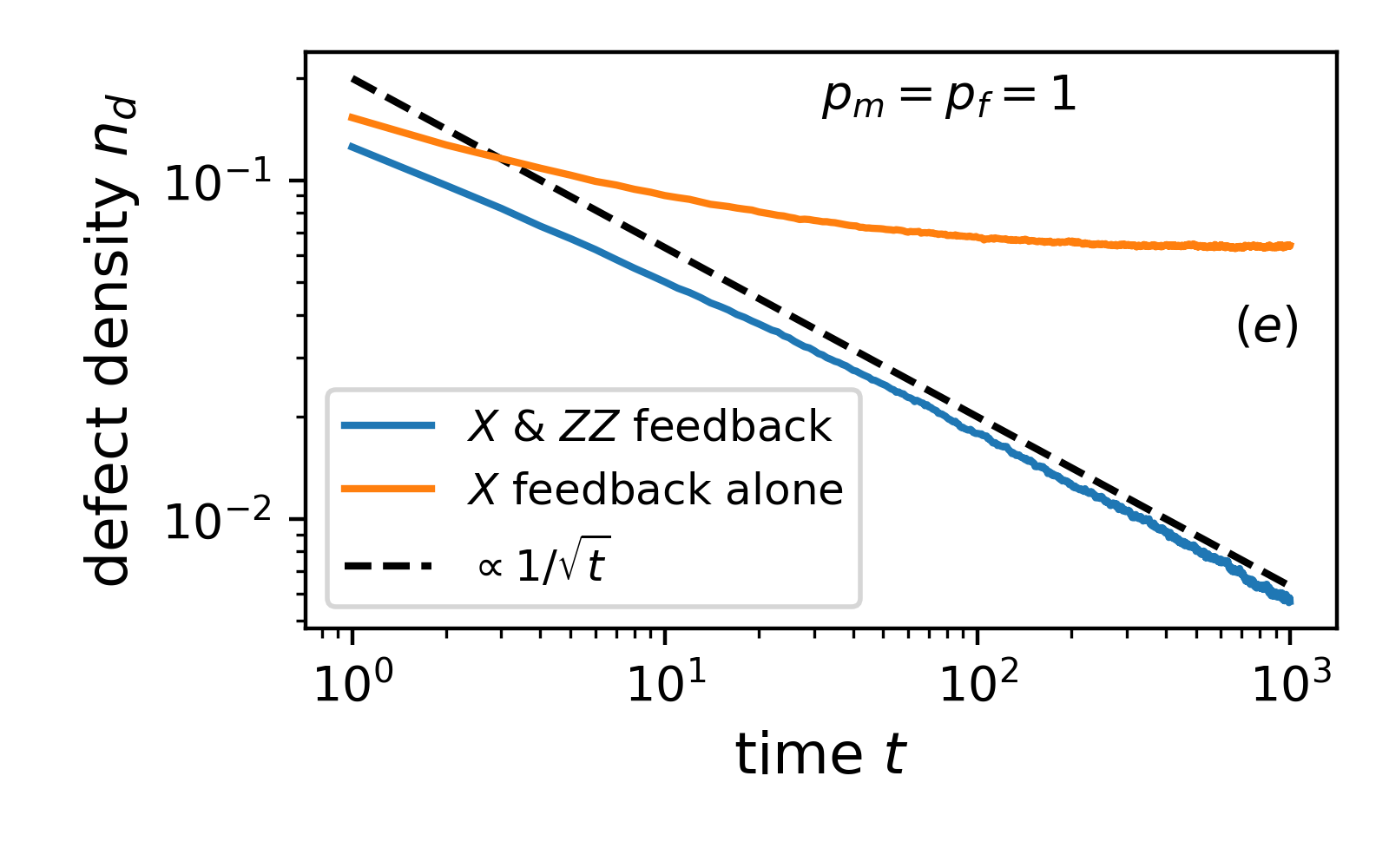}
\caption{Absorbing-state phase transition in the cluster-state models. The left (right) column corresponds to $Z$ ($X$ \& $ZZ$) corrective operations. (a)-(b) The running estimate of critical exponent $\delta$ as a function of inverse time $1/t$. The critical $p_c$ is estimated to be the value of $p$ for which the curve remains constant as $1/t \to 0$, and an estimate for $\delta$ is that constant value.  (c)-(d) Scaling collapse of $n_d$ data to the form of specified in the main text. The critical exponents are those of directed percolation (c) and the parity conserving universality class (d). (e) The dynamics of the averaged defect density $n_d$ at $\pf=\pme=1$ for $X$ corrections alone compared to $X$ and $ZZ$. $X$ corrections alone are not enough to drive the system into the absorbing cluster state, while $ZZ$ corrections diffusively pair up defects. \label{fig:ZXZ_scaling}}
\end{figure}

First, we demonstrate how the forms of the measurement and feedback change under this basis transformation. A measurement of $Z_{i-1} X_{i} Z_{i+1} = -1$ corresponds to a measurement of $Z_i = -1$ in the new basis. The cases of feedback in the original basis
\begin{itemize}
    \item $Z_i$
    \item Equal likelihood of $X_{i-1}$, $X_{i+1}$, $Z_{i-1}Z_i$, or $Z_{i}Z_{i+1}$
    \item Equal likelihood of $X_{i-1}$ or $X_{i+1}$
\end{itemize}
are now 
\begin{itemize}
    \item $X_i$
    \item Equal likelihood of $X_{i-2} Z_{i-1} X_i$, $X_i Z_{i+1} X_{i+2}$, $X_{i-1}X_{i}$, or $X_{i}X_{i+1}$
    \item Equal likelihood of $X_{i-2} Z_{i-1} X_i$ or $X_i Z_{i+1} X_{i+2}$
\end{itemize}
in the new basis.

From our discussion in Sec.~\ref{app:classical_polarized}, particularly that describing the action of Pauli strings, the classical stochastic process describing the diagonal of averaged density matrix inherits the following steps from the feedback above. Given that a site $i$ is $1$, during a measurement of that site, the respective steps below occur with probability $p_{\rm m} p_{\rm f}$:
\begin{itemize}
    \item Site $i$ is flipped
    \item Site $i$, along with a random choice of the four sites $i-2$, $i-1$, $i+1$, $i+2$, is flipped 
    \item Site $i$, along with a random choice of the two sites $i-2$ and $i+2$, is flipped.
\end{itemize}

The three-site random unitary used in all three models, written out in Eqs.~\ref{eq:ZXZ_unitary} and \ref{eq:ZXZ_unitary_basis}, transforms to a five-site operator under the basis transformation used above. On averaging in the doubled Hilbert space, the resulting operator once again does not mix the diagonal and off-diagonal elements of the averaged density matrix. The diagonal part of the density matrix transforms under a $32$ by $32$ matrix corresponding to the following classical stochastic process step: if the central bit is $0$, do nothing. If it's $1$, look at the parity of the surrounding four bits, two on each side, and sample those four bits with equal probability from the eight bitstrings consistent with that parity. For example, the state $00100$, with the operator centered on $1$, will with equal probability $1/8$ be sampled from the set $\{ 00100, 00111, 01101, 01110, 10101, 10110,11100, 111111\}$. Together with the feedback rule above, this fully specifies the classical stochastic process that describes the diagonal of the density matrix in this transformed basis. As the $Z_{i-1} X_i Z_{i+1}$ stabilizers become diagonal $Z_i$ in the transformed basis, we can find their expectation values through averaging over classical realizations of the stochastic process similarly to the polarized state.

\section{Dynamics for the cluster absorbing-state}

In this section, we discuss the critical properties and dynamics of the absorbing state transition in two of the models that absorb into the cluster state. We study the defect density, now defined as: $ n_d  = L^{-1}\sum_i (1 - Z_{i-1}X_iZ_{i+1})/2$. Our first model, with $Z$ feedback, does not preserve any symmetries, so it is natural to expect it to be in the directed percolation universality class. We show that this is indeed the case by showing consistency of the critical exponents with $\delta_{DP}$ and $\nu_{DP \parallel}$ in Fig.~\ref{fig:ZXZ_scaling}: In (a), we again estimate $p_c$ and $\delta$ by considering the separatrix; we find that $\delta$ is consistent with $\delta_{DP}=.159$ and we estimate $p_c = .1565(15)$. Pairing these values of $\delta_{DP}$ and $p_c$ with $\nu_{DP \parallel} = 1.73$ yields collapse of our data in (c).

On the other hand, our second model, with $X$ and $ZZ$ feedback, has a relevant symmetry: the parity of defect number is preserved. This symmetry is known to change the absorbing state transition to the parity conserving or PC class~\cite{Hinrichsen_2000}. Indeed, in estimating $\delta$ from the separatrix in Fig.~\ref{fig:ZXZ_scaling}(b), we see that it is consistent with $\delta_{PC} =.286$ and inconsistent with $\delta_{DP}$. The resulting $p_c=.606(8)$. We also see collapse of these curves in (d) when using $\nu_{PC ||} = 3.22$. Thus, we see that the addition of the extra symmetry indeed modified the universality class to the PC class. 

Finally, in Fig.~\ref{fig:ZXZ_scaling}(e) we tune to the point $\pme=\pf=1$, which is the point of highest feedback, and show that the model with only $X$ corrections in its feedback does not converge to the target state, while the model with both $X$ and $ZZ$ does. In the latter case, as noted in the main text, the defect density goes down diffusively ($n_d \propto t^{-1/2}$) as random-walking defects are annihilated in pairs in such parity-conserving models.

% using bibtex
\bibliography{main}